\documentclass[graybox]{SNmult}

\usepackage{type1cm}        
\usepackage{makeidx}         
\usepackage{graphicx}        
\usepackage{multicol}        
\usepackage[bottom]{footmisc}

\usepackage{newtxtext}       %
\usepackage{nicefrac}       %
\usepackage[varvw]{newtxmath}       

\makeindex             

\def\1{\mathbf{1}}

\def\cB {\mathscr{B}}
\def\cC{\mathcal{C}}

\def\cF{\mathscr{F}}

\def\cI{\mathscr{I}}

\def\RR{{\mathbb R}}

\def\EE{{\mathbb E}}

\def \0{{\mathbf{0}}}

\def \vecw {\mathbf{w}}

\def \Pr {\mathrm{Pr}}

\def \d {\mathrm{d}}
\def \vecc {\mathbf{c}}

\def \mymid {\,:\,}

\begin{document}

\title*{Cooperative games with uncertainty: a survey}
\author{Michel Grabisch\thanks{Michel Grabisch acknowledges the financial support
by the European Union HORIZON-WIDERA-2023-TALENTS-01 project No 101183743
(AGATE).}\orcidID{0000-0002-3283-1496} and\\ Silvia Lorenzini\orcidID{0009-0008-1785-3065}}
\institute{Michel Grabisch\at Charles University, Prague, Czech Republic, and Universit\'e Paris
  1 Panth\'eon-Sorbonne, Paris, France. \email{michel.grabisch@univ-paris1.fr}
\and Silvia Lorenzini \at Universit\`a degli Studi di Perugia, Perugia, Italy. \email{silvia,lorenzini@dottorandi.unipg.it}}
%
%
\maketitle
\abstract*{The main stream of the literature in cooperative game theory considers
  games as being deterministic. In this survey, we review the main models
  incorporating uncertainty in cooperative games, starting from the seminal
  paper of Charnes and Granot (1976) and finishing with our own contribution,
  called Bel coalitional games. The different models are explained and compared,
  making a focus on the notion of allocation and core, since all the described
  models propose a notion of core. 
\keywords{Cooperative game $\cdot$ Uncertainty $\cdot$ Core $\cdot$ Nucleolus
  $\cdot$ Bargaining set}}

\abstract{The main stream of the literature in cooperative game theory considers
  games as being deterministic. In this survey, we review the main models
  incorporating uncertainty in cooperative games, starting from the seminal
  paper of Charnes and Granot (1976) and finishing with our own contribution,
  called Bel coalitional games. The different models are explained and compared,
  making a focus on the notion of allocation and core, since all the described
  models propose a notion of core. 
\keywords{Cooperative game $\cdot$ Uncertainty $\cdot$ Core $\cdot$ Nucleolus
  $\cdot$ Bargaining set}}

\section{Introduction}\label{sec:intro}
Cooperative games have a long history since their inception in the 40s by von
Neumann and Morgenstern \cite{vnm44}. Many variants have been proposed around
the two main types of cooperative games, namely games with transferable utility
(TU) and games with non-transferable utility (NTU), but the literature in the
recent years has more focused on TU-games. Essentially, a TU-game assigns a real
number $v(S)$ to a coalition $S$ of players who have decided to form an agreement, and this
number quantifies the benefit of their cooperation. This very simple model has
been generalized in many directions, among which we may cite (1) games with
restricted cooperation, where the collection of coalitions which may form is not
the whole power set but some subcollection having specific properties (closed under
union, or intersection, etc.)(see, e.g., \cite{gra12a}); (2) games in partition
function form \cite{thlu63}, where the quantity $v(S)$ depends on the arrangement
of the remaining players in $N\setminus S$; (3) multichoice games and fuzzy games
(see, e.g., \cite{hsra93,brditi05}), where players have a finite (multichoice)
or continuous (fuzzy) set of levels of participation to the game, instead of a
binary choice (participate, do not participate).

However, in any of these variants, the quantity assigned to a coalition is always a real
number, meaning that under a given configuration, the benefit of cooperation is
perfectly determined.  Obviously, this may not be the case in practice, for
essentially two reasons. The first one is that the quantity may be imprecisely
known, and then $v(S)$ could be an interval or a fuzzy number. These games are
usually called imprecise games, or interval-valued games, or fuzzy-valued games,
see, e.g., \cite{brbrgoti10,mar94}. The second one is that the quantity may
depend on the state of the world, which is unknown, therefore $v(S)$ is
uncertain and could be seen for example as a random variable. This survey
precisely addresses this case. The literature on ``uncertain'' games is
relatively small and is not so well known. This survey will show that the topic
is far to be simple and very different models exist, which are of high interest,
with challenging questions. Our hope is that by this survey we could contribute to a revival of
this topic and to the dissemination of this literature.

In Section~\ref{sec:prel} we introduce the necessary background on cooperative
games. Then, from Section~\ref{sec:chgr} to \ref{sec:hahe}, we detail the main
models of the past literature, and in Section~\ref{sec:bel} we describe our
contribution to the field with the notion of Bel coalitional games.

\section{Preliminaries}\label{sec:prel}
(see, e.g., \cite{masoza13,pesu03,pet08} for a extensive exposition) A {\it game
  in characteristic form with transferable utility} (abbreviated hereafter by
        {\it TU-game} or {\it game}) is a pair $(N,v)$ where $N=\{1,\ldots,n\}$
        is a finite set of players or agents, and $v:2^N\rightarrow \RR$ is a
        mapping assigning to every {\it coalition} $S\subseteq N$ (i.e., a
        subset of players) a number $v(S)$, representing the {\it worth} of
        coalition $S$. Coalition $N$ is called the {\it grand coalition} and it is assumed
        that $v(\varnothing)=0$. In most applications, the worth of $S$ is the
        benefit earned by coalition $S$ due to the cooperation of its
        members. It can also be interpreted as a cost of using a certain
        service. In the sequel, we will stick to the first interpretation.

A game $(N,v)$ is {\it superadditive} if $v(S\cup T)\geqslant v(S)+v(T)$ for
every disjoint coalitions $S,T$. It is {\it convex} if the following
inequalities hold for every $S,T\subseteq N$:
\[
v(S\cup T) + v(S\cap T)\geqslant v(S) + v(T).
\]
Convexity implies superadditivity. Subadditivity and concavity are defined
likewise with the reversed inequality.

        One of cooperative game theory's aim is to solve the problem of sharing the
worth of the grand coalition $v(N)$ between all players in a rational and fair
way. Hence, a {\it solution} in this framework is a systematic way of dividing
$v(N)$. A {\it payoff vector} $x=(x_1,\ldots,x_n)$ is a vector in $\mathbb{R}^N$
defining the payoff given to any agent in $N$. A payoff vector $x$ is
\begin{itemize}
\item {\it feasible} if $\sum_{i\in N} x_i \leqslant v(N)$;
\item {\it efficient} if $\sum_{i\in N} x_i = v(N)$;
\item {\it individually rational} if $x_i \geqslant v(\{i\})$, for every player $i\in N$.
\end{itemize}
We denote by $X(N,v)$ the set of {\it pre-imputations}, that is the set of all
efficient payoff vectors, while $I(N,v)$ denotes the set of {\it imputations},
that is, individually rational pre-imputations. The main solution concept for cooperative games is the {\it core}.
\begin{definition}
The \textit{core} of a game $(N,v)$ is defined by 
    $$\mathcal{C}(N,v)=\{x\in X(N,v): x(S) \geqslant v(S),\,\forall\, S\subseteq N\},$$
    where $x(S)=\sum_{i\in S} x_i$, for all $S\subseteq N$.
\end{definition}
For a given game, there may be no payoff vector that satisfies the stated condition, i.e., the core may be empty. The analysis of the core in cooperative games is primarily concerned with establishing a condition that characterizes the nonemptiness of the core. Using the notion of balanced games, the Bondareva-Shapley theorem gives necessary and sufficient conditions for nonemptiness of the core.

A collection $\cB$ of nonempty subsets of $N$ is {\it balanced} if there exists
positive coefficients $(\lambda_S)_{S\in\cB}$ (called {\it balancing weights}) such that $\sum_{S\in\cB,S\ni
  i}\lambda_S = 1$, for all $i\in N$. A balanced collection is {\it minimal} if
no proper subcollection is balanced (equivalently, if it has a unique system of
balancing weights). 

A game $(N,v)$ is {\it balanced} if for every balanced collection $\cB$ it holds
\[ 
\sum_{S\in\cB} \lambda_Sv(S)\leqslant v(N).
\]
It is well-known that a game is balanced if and only if the above inequality is
satisfied for every {\it minimal} balanced collection.
\begin{theorem}
(Bondareva \cite{bon63}, Shapley \cite{sha67}) A game $(N,v)$ has a nonempty core if and only if it is balanced.
\end{theorem}

A well-known result is that any convex game has a nonempty core. More precisely,
the core of a convex game contains all marginal vectors, which happen to
coincide with the set of vertices of the core (the converse is also true). Given
a game $(N,v)$ and a permutation $\sigma$ on $N$, the {\it marginal vector}
$m^{\sigma,v}$ is a payoff vector in $\RR^N$ defined by:
\[
m^{\sigma,v}_{\sigma(i)} = v(S_i)-v(S_{i-1}), \quad i=1,\ldots,n
\]
where $S_i=\{\sigma(1),\ldots,\sigma(i)\}$.

The fact that the core can be empty or very large motivates the study of alternative solution concepts. The nucleolus \cite{sch69} is a solution concept defined through the notion of excess. The {\it excess} of some coalition $S$ at $x$ is given by
$$e(S,x)=v(S)-x(S)$$
and it provides a measure of the dissatisfaction of coalition $S$ with respect to the payoff vector $x$.

Then, we introduce the vector of decreasing excesses
$$\theta(x)=(e(S_1,x),\dots,e(S_{2^N},x))$$
where $S_i$, $i=1,\dots,2^N$, are all the coalitions indexed such that
$$e(S_1,x)\geqslant\cdots\geqslant e(S_{2^N},x).$$ 

The lexicographic order is used to compare the vectors $\theta(x)$ and
$\theta(y)$, denoted by $\succsim_L$. Given two vectors $\theta(x)$ and
$\theta(y)$, we write
$\theta(x)\succsim_L\theta(y)$ if either
$\theta(x)=\theta(y)$ or there exists $k$,
$1\leqslant k\le N$, such that $\theta_k(x)>\theta_k(y)$ and
$\theta_i(x)=\theta_i(y)$, for all $i$ such that $1\le
i<k$. In particular we have that
$\theta(x)\succ_L\theta(y)$ if
$\theta(x)\succsim_L \theta(y)$ and
$\theta(y)\not\succsim_L\theta(x)$.

\begin{definition}
Let $X\subset \RR^{N}$. The {nucleolus w.r.t. $X$} of a game $(N,v)$ is defined by
$$\mathcal{N}(N,v,X)=\{x\in X \,:\, \theta(x)\precsim_L\theta(y),\, \forall x\in X\}.$$
\end{definition}

The {\it pre-nucleolus} $\mathcal{PN}(N,v)$ is defined as the nucleolus w.r.t. the set of all efficient vectors $x\in\RR^{N}$. It is possible to prove that the nucleolus relative to any compact set is nonempty and if the set is also convex, then the nucleolus relative to that set consists of a single vector. Then, as a consequence, the pre-nucleolus of a coalitional game is a nonempty compact set.

We omit for brevity the introduction of other well-known solution concepts like
 the kernel, and conclude this section with the introduction of
the {\it bargaining set}, as we will generalize this notion for games with uncertainty. Several definitions of the bargaining set have
been proposed in the literature. We present here two of the most widely used
ones, namely those of Davis and Maschler \cite{dama67} and Mas-Colell
\cite{mas89}. In the Davis–Maschler definition, given a payoff vector $x$, an
agent $i$ raises an {\it objection} against another agent $j$ by proposing a
payoff $y$ for a coalition $S$ that includes $i$ but excludes $j$. The payoff
$y$ must improve the payoff of all members of $S$ relative to $x$. The objection
is considered {\it justified} if $j$ cannot propose an objection against
$i$. The main difference with the Mas-Colell’s definition is that objections are
formulated by one coalition against another, rather than by a single player
against another.

\begin{definition}
The {\it bargaining set} of a game $(N,v)$ is defined by 
$$\mathcal{B}(N,v)=\{x\in I(N,v)\,:\,\not\exists y\,\mbox{which is a justified objection to}\, x\},$$
i.e., no agent has a justified objection against any other agent in his coalition.
\end{definition}

It is shown that the bargaining set is never empty since it contains the
pre-nucleolus, and is a compact set.
Furthermore, since no objection can arise at a payoff in the core we have that, in general,
$$\mathcal{C}(N,v)\subseteq\mathcal{B}(N,v)$$
while the equality holds for convex games (when the Davis-Maschler definition is
used).

\section{Cooperative games in stochastic characteristic form}\label{sec:chgr}
The first attempt to introduce uncertainty in cooperative games is due to
Charnes and Granot in a series of papers \cite{chgr73,chgr76,chgr77,gra77}. The
general idea behind these works is the following: $v(S)$ is no more a
deterministic quantity but a random variable with known (cumulative)
distribution function $F_S$. The main question is how to define a payment to the
players without knowing with certainty the worth of $S$. Charnes and Granot
define a payoff process in two steps:
\begin{enumerate}
\item Step 1: before observation (realization of $v(S)$ for all coalitions $S$),
  the coalitions are promising prior payoffs to their members, in order that
  there will be a good chance to satisfy the promises;
\item Step 2: after observation, in case of impossibility to pay the members the
  promised payoffs, a modification of the payoffs occurs, so that it will raise
  the least objections among players. 
\end{enumerate}

We introduce the model formally. Let $N=\{1,\ldots,n\}$ the set of players, and
consider for each coalition $S\neq\varnothing$ a random variable $v(S)$ taking
nonnegative values. This defines $(N,V)$, a {\it chance-constrained game}, $V$
denoting the set of random variables $v(S)$ with distribution $F_S$, for all
coalitions $S$. We
suppose that the grand coalition $N$ forms. A promised prior payoff for $N$ is a vector
$x=(x_1,\ldots,x_n)\in\RR^N_+$ which satisfies the following condition:
\begin{equation}\label{eq:Y}
x(N) = F_N^{-1}(\alpha(N))=\min\{t\mymid F_N(t)\geqslant \alpha(N)\},
\end{equation}
where $\alpha(N)\in\left]0,1\right[$ is the fractile associated to the grand
    coalition. Note that if $\alpha(N)=\frac{1}{2}$, $x(N)$ is the median of the
    distribution of $v(N)$. This condition defines the set of feasible prior
    payoffs, denoted by $Y$.

    \medskip
    
In \cite{chgr76}, Charnes and Granot define the prior nucleolus, the term
``prior'' referring to Step 1 of the payoff process. For a given feasible prior
payoff $x\in Y$ and a coalition $S$, the {\it probabilistic excess} of $S$ is
defined by $e(S,x) = 1-F_S(x(S))$ (recall that $x(S)=\sum_{i\in S}x_i$). Clearly, the
greater $e(S,x)$, the greater the dissatisfaction of receiving $x(S)$, as in the classical
definition. Based on this interpretation, we say that for two feasible prior
payoffs $x,z\in Y$, $x$ is {\it more acceptable} than $z$ if
\[
\max_{S\subseteq N} e(S,x)<\max_{S\subseteq N} e(S,z).
\]
As in the classical case, the nucleolus is just a refinement of acceptability
using the lexicographic order. Define $K(x)\in\RR^{2^n-1}$ the vector of excesses
in decreasing order. Then the {\it prior nucleolus} is defined by
\[
N_1(N,V) = \{x\in Y\mymid K(x)\precsim_{\rm lex}K(y),\forall y\in Y\}.
\]
It is proved in \cite{chgr76} that the prior nucleolus is always nonempty. In
addition, the following holds.
\begin{theorem}
If $F_S$, $S\subseteq N$, are strictly monotone increasing, then $N_1(N,V)$ is a
singleton for a possibly restricted game to $S_1,\ldots, S_k$, s.t.,
$\{1^{S_1},\ldots, 1^{S_k}\}$ contains a basis of $\RR^n$.
\end{theorem}

\medskip

In \cite{gra77}, Granot extends their previous work by 
considering games with coalition structures, redefining the prior nucleolus in
this more general framework, and proposing the notions of prior kernel and prior
bargaining set. A {\it coalition structure} is a collection
$\cB=\{B_1,\ldots,B_m\}$ of coalitions which forms a partition of $N$. It
indicates a prior grouping of the players (a classical TU-game corresponds to
$\cB=\{N\}$). The definition of the set of feasible prior payoffs (called now
{\it prior individual rational payoff configuration}, abbreviated by p.i.r.p.c.)
is simply an adaptation of (\ref{eq:Y}) to the coalition structure. For
$x\in\RR^N_+$ and $\alpha=(\alpha(B_1),\ldots,\alpha(B_m))\in\left]0,1\right[^m$, the triple
    $(x;\cB;\alpha)$ is a p.i.r.p.c. if
    \[
x(B_j) =  F_{B_j}^{-1}(\alpha(B_j))=\min\{t\mymid F_{B_j}(t)\geqslant \alpha(B_j)\}
    \]
    for all $j=1,\ldots,m$, where $\alpha(B_j)$ is the fractile associated to
    $B_j$.  Granot proposes a definition of the prior kernel and the prior
    bargaining set. For the latter, although it is defined like in the classical
    definition of Davis and Maschler \cite{dama67} via objections and
    counterobjections, there are fundamental differences. In particular, given
    $(x;\cB;\alpha)$ a p.i.r.p.c., an objection of a player $i$ against a player
    $j$, both being in the same coalition $B_k$, is defined as another coalition
    $G$ containing $i$ but not $j$, such that the probability that player $i$
    will realize his payoff $x_i$ is greater in $G$ than in $B_k$, and similarly
    for the other players in $G$. In the classical definition, an objection is
    another payoff vector $y$ and a coalition $G$ such that $y$ is affordable
    for $G$ and strictly better than $x$ for the members of $G$.

    \medskip

Finally, we address the second step of the two-stage payoff process. There is an
easy way to deal with this second step given in \cite{gra77}. Let $x\in\RR^N$ be
the prior solution and let $v$ denotes the realization of $(N,v)$. We are
looking for a payoff vector $y\in\RR^N$ being nonnegative and satisfying the
condition $y(B_k)=v(B_k)$ for all $k=1,\ldots, m$. In order to be as close as
possible to $x$, Granot suggests to take the unique $y$ satisfying the above condition and
preserving the ratio $x_i/x_j$ for all $i,j\in B_k$, i.e., $x_i/x_j=y_i/y_j$. A
more sophisticated way is proposed in \cite{chgr77} tailored for the nucleolus.

\section{Cooperative games with stochastic payoffs}\label{sec:subo}
About twenty years later, Suijs, Borm and colleagues revisit the work of Charnes
and Granot in two compagnion papers \cite{subowati99,subo99}. While in the model
of Charnes and Granot, risk neutrality is assumed for the players, Suijs et al.
allow for a more general preference model. Also, they introduce actions to
choose for the players, resulting in a more complex model.

A {\it game with stochastic payoff} is a tuple $\Gamma=(N,(A_S)_{S\subseteq
  N},(X_S)_{S\subseteq N},(\succcurlyeq)_{i\in N})$, where $A_S$ is a set of
actions for coalition $S$, $X_S$ is a mapping from $A_S$ to $L^1(\RR)$, the set
of all real-valued random variables with finite expectation, representing the
random benefit of coalition $S$ (classically $v(S)$) when action $a$ is chosen,
and $\succcurlyeq_i$ is the preference of player $i$ over $L^1(\RR)$.  The set
of games with stochastic payoffs on $N$ is denoted by $SG(N)$.

Compared with the model of Charnes and Granot, in the latter model there is no
choice of action, and the preference is expressed via the distribution function
of $v(S)$ for all $S$. 

Suijs et al. define payoff vectors for coalitions as follows. For a coalition
$S$ and an action $a\in A_S$, a pair $(d,r)\in \RR^S\times \RR^S_+$ is an {\it allocation} if
\begin{enumerate}
\item $\sum_{i\in S} d_i = \EE(X_S(a))$ (i.e., $d$ distributes the expected
  value of $X_S(a)$ to the members of $S$);
\item $\sum_{i\in S}r_i=1$ (sharing vector).
\end{enumerate}
Then, the corresponding payoff given to player $i\in S$, when action $a$ is
chosen and $X_S(a)$ is known, is defined by:
\[
(d,r|a)_i = d_i + r_i(X_S(a) - \EE(X_s(a))
\]
The first term is the allocation of the ``expectation of $v(S)$'' and the second
term is the allocation of the residual, called the risk of $X_S(a)$. We denote
by $Z(S)$ the set of all 
allocations for $S$.

The {\it core} of a game $\Gamma$ in $SG(N)$ is defined as the set of
allocations on $N$ which no coalition $S$ can block, formally:
\[
\cC(\Gamma) = \{(d,r|a)\in Z(N)\mymid \nexists S,(\hat{d},\hat{r}|\hat{a})\in
Z(S) \text{ s.t. }  (\hat{d},\hat{r}|\hat{a})_i\succ_i(d,r|a)_i,\forall i\in S\}
\]

\medskip

Suijs et al. consider then a particular form for the preference over $L^1(\RR)$,
which permits to get a characterization of nonemptiness of the core similar to
the Bondareva-Shapley theorem.

Let $X,Y\in L^1(\RR)$ with distributions $F_X,F_Y$. For a given
$\alpha\in\left]0,1\right[$, $X$ is $\alpha$-preferred to $Y$ (denoted by
    $X\succcurlyeq_\alpha Y$) if $u^X_\alpha\geqslant u^Y_\alpha$, where
    $u^X_\alpha:=\sup\{t\mymid F_X(t)\leqslant \alpha\}$ is the
    $\alpha$-quantile of $X$ (e.g., $\alpha=\nicefrac{1}{2}$ yields the median
    value). A game where the preference of agent $i$ is given by
    $\succcurlyeq_{\alpha_i}$ for all $i\in N$ is denoted by $\Gamma_\alpha$,
    with $\alpha=(\alpha_1,\ldots,\alpha_n)$.

The greater the value of $\alpha$, the more risk-loving is the preference
$\succcurlyeq_\alpha$, in the sense that there are more random variables $X\in
L^1(\RR)$ preferred to the certainty equivalent $\EE(X)$.

A game $\Gamma_\alpha$ is said to be {\it balanced} if for every balanced
collection $\cB$ with balancing weights $(\lambda_S)_{S\in\cB}$ it holds:
\[
\sum_{S\in\cB}\lambda_S\max_{a\in A_S}\max_{i\in
  S}u_{\alpha_i}^{X_S(a)}\leqslant \max_{a\in A_N}\max_{i\in N}u_{\alpha_i}^{X_N(a)}.
\]
Observe that in both sides, the highest quantile among the players is chosen,
which means the most risk-loving behavior. If the game becomes deterministic,
then $u^{X_S(a)}_{\alpha_i}$ simply reduces to $v(S)$, and
$u^{X_N(a)}_{\alpha_i}$ reduces to $v(N)$, so that we recover the classical
definition of balanced games. The main result of \cite{subowati99} is the
following:
\begin{theorem}
The core of a game $\Gamma_\alpha$ is nonempty if and only if $\Gamma_\alpha$ is
balanced.
\end{theorem}
What is instrumental in the proof is the following: an allocation $(d,r|a)\in
Z(N)$ belongs to the core $\cC(\Gamma_\alpha)$ if and only if it satisfies for
every coalition $S$:
\[
\sum_{i\in S}\Big(d_i+r_i(u^{X_N(a)}_{\alpha_i} - \EE(X_N(a))\Big)\geqslant
  \max_{a'\in A_S}\max_{i\in S}u^{X_A(a')}_{\alpha_i}.
  \]
  Keeping in mind that $\max_{a'\in A_S}\max_{i\in S}u^{X_A(a')}_{\alpha_i}$
  plays the role of $v(S)$, we retrieve a relation similar to the classical
  definition of the core, where the payoff to coalition $S$ is defined using the
  $\alpha_i$-quantiles. In particular, for a deterministic game, we get exactly
  the definition of the core.

  \medskip

  Suijs et al. consider in the last part of \cite{subowati99} another type of
  preference among random variables, which is stochastic dominance: $X$ {\it
    stochastically dominates} $Y$ (at first order) (denoted by $X\succ_F
  Y$) if $F_X(t)\leqslant F_Y(t)$ 
  for all $t$. This order is largely incomplete, and since $X\succ_F Y$
  implies $X\succcurlyeq_\alpha Y$, we get that if $\cC(\Gamma_\alpha)\subseteq
  \cC(\Gamma_F)$, where $\Gamma_F$ is the game $\Gamma_\alpha$ whose preference
  has been replaced by $\succ_F$. 

  \medskip

  In \cite{subo99}, Suijs and Borm simplify their setting by considering there
  is only one action to choose for each coalition $S$, and
  $\Pr(X_\varnothing(a)=0)=1$. In addition, the preference $\succcurlyeq_i$ over
  random variables in $L^1(\RR)$ is supposed to be complete, transitive and
  continuous (i.e., the sets $\{Y\in L^1(\RR)\mymid Y\succcurlyeq X\}$ and
  $\{Y\in L^1(\RR)\mymid Y\preccurlyeq X\}$ are closed). An allocation
  for coalition $S$ is a pair  $(d,r)\in\RR^S\times \RR^S$ such that $r$ is a
  sharing vector ($r_i\geqslant 0$, $\sum_{i\in S}r_i=1$) and $\sum_{i\in
    S}d_i\leqslant 0$. The corresponding (random) payoff to player $i$ is
  $d_i+r_iX_S$ (observe the difference with definition above). The set of
  allocations for $S$ is denoted by $Z(S)$, while $IR(S):=\{(d,r)\in Z(S)\mymid
  d_1+r_iX_S\succcurlyeq_i X_{\{i\}}\}$ is the set of individually rational
  allocations.

  A game $\Gamma$ is said to be {\it superadditive} if for every disjoint
  coalitions $S,T$, for every $(d^S,r^S)\in Z(S), (d^T,r^T)\in Z(T)$, there
  exist $(d^{S\cup T},r^{S\cup T})\in Z(S\cup T)$ such that
  \begin{align*}
    d_i^{S\cup T}+r_i^{S\cup T}X_{S\cup T} & \succcurlyeq_i d_i^S+r_i^SX_S,
    \forall i\in S\\
    d_i^{S\cup T}+r_i^{S\cup T}X_{S\cup T} & \succcurlyeq_i d_i^T+r_i^TX_T,
    \forall i\in T.
  \end{align*}
Comparing this definition with the definition for classical TU-games shows
essential differences, first because the property is expressed in terms of
(random) payoffs instead of the characteristic function $v$, and second because
no (super)additivity explicitly appears, but rather a monotonicity
property. However, the definition can be related to the definition of
superadditivity for NTU-games. We recall that a NTU (non-tranferable utility)
game assigns to every coalition $S$ a set of feasible payoffs $V(S)\subseteq
\RR^S$. An NTU-game $V$ is said to be superadditive if $V(S)\times V(T)\subseteq
v(S\cup T)$, which expresses the fact that any feasible payoff for $S$ combined
with any feasible payoff for $T$ is feasible for $S\cup T$. This is exactly the
situation described by the superadditivity of $\Gamma$; whatever the payoffs
chosen for $S$ and for $T$, there exists a payoff for $S\cup T$ which is at
least as preferred as them for all players in $S\cup T$.

Convexity is defined using a property for TU-games which is equivalent to the
usual definition: for every $S,T$ such that $S\subseteq T$ and every $U\subseteq
N\setminus T$,
\[
v(S\cup U)-v(S) \leqslant v(T\cup U)-v(T).
\]
A stochastic game $\Gamma$ is {\it convex} if for all $S,T,U$ as above, for all
$(d^S,r^S)\in IR(S)$, $(d^T,r^T)\in IR(T)$, and $(d^{S\cup U},r^{S\cup U})\in
Z(S\cup U)$ satisfying for all $i\in S$
\[
d_i^{S\cup U}+r_i^{S\cup U}X_{S\cup U}\succcurlyeq_i d_i^S+r^S_iX_S,
\]
there exists an allocation $(d^{T\cup U},r^{T\cup U})\in Z(T\cup U)$ such that
\begin{align*}
  d_i^{T\cup U}+r_i^{T\cup U}X_{T\cup U} & \succcurlyeq_i d_i^T+r_i^TX_T,
  \forall i\in T\\
  d_i^{T\cup U}+r_i^{T\cup U}X_{T\cup U} & \succcurlyeq_i d_i^{S\cup
    U}+r_i^{S\cup U}X_{S\cup U},   \forall i\in U.  
\end{align*}
Based on this definition, Suijs and Borm show that the classical result of
nonemptiness of the core for convex games still holds in this context.
\begin{theorem}
Let $\Gamma$ be a stochastic cooperative game. If $\Gamma$ is convex, then
$\cC(\Gamma)\neq\varnothing$. 
\end{theorem}
The proof is based on the construction of a particular allocation, similar to
the marginal vectors of classical TU-games. It is proved that any such
allocation belongs to the core when the game is convex. However, contrarily to
classical TU-games, the fact that all such marginal vectors belong to the core,
the game is not necessarily convex. 

\section{Uncertain cooperative games}\label{sec:boss}
The next attempt to introduce uncertainty in cooperative games seems to have
been proposed by Bossert et al. in 2005 \cite{bodepe05}, under the name of
uncertain cooperative games, meaning that the worth of a coalition depends on
the (unknown) state of nature. Most of the exposition in \cite{bodepe05} is done
with only two states of nature, a way which we follow here also. Let $N$ be a
set of $n$ players, and $\cF$ be a subcollection of
$2^N\setminus\{\varnothing,N\}$. We deal throughout this section with TU-games
defined on $\cF\cup\{N\}$. A collection $\cB\subseteq \cF$ is said to be {\it
  weakly balanced} if it contains a balanced collection.

Let us consider two classical TU-games $w,w'$, which will be seen as the
uncertain game with two states of nature. Take two pre-imputations $q,z$ for the
game $w$, and two pre-imputations $q',z'$ for the game $w'$. We should interpret
$(q,q')$ and $(z,z')$ as two uncertain allocations for the uncertain game.  
We say that $(q,q')$ {\it dominates} $(z,z')$ if for all $S\in\cF$,
\[
\min\{q(S)-w(S),q'(S)-w'(S)\}>\min\{z(S)-w(S), z'(S)-w'(S)\}.
\]
In words,  $(q,q')$ dominates $(z,z')$ if for each coalition, the corresponding
excess vector for $(q,q')$  is better than the excess vector for $(z,z')$, in
the worst state of nature.

The main result of the paper is to characterize undominated allocations. This is
done through the following equivalence: letting $v:=w-w'$ and $x:=z-z'$, $x$ (or
$(z,z')$) is undominated if and only if for every side payment $y,y'$ (i.e.,
$y(N)=y'(N)=0$) the following implication holds for all $S\in\cF$:
\begin{multline*}
\min\{x(S)+y(S),v(S)+y'(S)\}\geqslant\min\{x(S),v(S)\} \Rightarrow\\ \min \{x(S)+y(S),v(S)+y'(S)\}\not>\min\{x(S),v(S)\} .
\end{multline*}
Using the above notation, the following holds.
\begin{theorem}
$x$ is undominated if and only if $\{S\in \cF\mymid x(S)\geqslant v(S)\}$ or
  $\{S\in\cF\mymid x(S)\leqslant v(S)\}$ is weakly balanced. 
\end{theorem}
Observe how the analysis on an uncertain game and uncertain allocation is turned
into the analysis on a classical game and allocation.  
  
 Recall that a pre-imputation $x$ is in the core of a TU-game $v$ if
 $x(S)\geqslant v(S)$ for all $S\in\cF$. The anti-core is defined in a symmetric
 way: $x$ in in the anti-core of $v$ if $x(S)\leqslant v(S)$ for all
 $S\in\cF$. A pre-imputation is said to be {\it stable} for $v$ if it is either in the
 core or in the anti-core  of $v$. It follows immediately from the above
 theorem  and the Bondareva-Shapley theorem that, when $\cF$ is balanced, any stable
 allocation $x$ is undominated.

\section{Bayesian coalitional games}\label{sec:iesh}
The next notable model of uncertain game in chronological order seems to be the
one proposed by Ieong and Shoham, under the name of Bayesian coalitional games
\cite{iesh08}. Interestingly, this work does not emanate from the domain of game
theory like any other work cited in this survey, but comes from the field of
Artificial Intelligence (AI). By the way, the approach of the authors is more rooted
in Baysian decision theory than in game theory, and it offers a more general
framework than the one of Suijs et al. (see Section~\ref{sec:subo}), provided only
one action is available for each coalition. The authors cite other similar works
in the field of AI.

A {\it Bayesian coalitional game} is a tuple $(N,\Omega,P, (\cI_j)_{j\in
  N},(\succcurlyeq_j)_{j\in N})$ where $N=\{1,\ldots,n\}$ is the set of agents,
$\Omega=\{\omega_1,\ldots,\omega_m\}$ is the set of states of the world (and to
each $\omega\in\Omega$ a coalitional game $v_\omega$ is associated), $P$ is a
prior probability distribution on $\Omega$, common to all agents, and for each
agent $j$, $\cI_j$ is a partition of $\Omega$ and $\succcurlyeq_j$ is a
preference relation over the set $\RR^\Omega$ of payoff distributions. The
partition $\cI_j$ is called the {\it information partition}, which is public
knowledge, and it represents the equivalence classes of the indistinguishability
relation of agent $j$, i.e., two states of the world $\omega,\omega'$ in $I\in
\cI_j$ are indistinguishable for this agent.

Before uncertainty is resolved (the ex-ante situation), the exact worth of a
coalition being unknown, a payoff to an agent cannot be defined as a single
number, but rather by a {\it distribution} over $\Omega$. Ieong and Shoham
assume that agents enter into agreement {\it ex-ante} about how to divide the
worth of coalitions, which are called contracts. Once the uncertainty is
resolved (ex-post situation), each agent receives the amount specified by the
contract in the true state of the world. Formally, a {\it contract} among agents
of coalition $S$ (or {\it $S$-contract}) is a mapping
$\vecc^S:\Omega\rightarrow\RR^S$. A contract is {\it feasible} if $\sum_{j\in
  S}\vecc^S_j(\omega)\leqslant v_\omega(S)$ for all $\omega\in\Omega$. In what
follows, only feasible contracts are considered.

For two $S$-contracts $\vecc^S,\vecc'^S$, we write
$\vecc^S\succcurlyeq_S\vecc'^S$ if for every agent $j\in S$ it holds
$\vecc^S_j\succcurlyeq_j\vecc'^S_j$, where the subindex $j$ indicates the payoff
for agent $j$. We write $\vecc^S\succ_S\vecc'^S$ if preference is strict for all
agents in $S$.

\medskip

The main question addressed by Ieong and Shoham is the study of the core,
defined as the set of all $N$-contracts (called {\it grand contracts}) which are
not ``blocked'' by other coalitions. The notion of blocking depends on which
situation is considered: ex-ante, ex-interim and ex-post, leading to the
definitions of {\it ex-ante core}, {\it ex-post core} and {\it ex-interim core}. In the ex-ante
situation, a grand contract $\vecc^N$ is {\it ex-ante blocked} by a coalition
$S$ if there exists a $S$-contract $\vecc^S$ such that
$\vecc^S\succ_S\vecc^N_S$, where $\vecc^N_S$ indicates the restriction of
$\vecc^N$ to agents in $S$. In the ex-post situation, a grand contract $\vecc^N$
is {\it ex-post blocked} by a coalition $S$ if there exists $\omega^*\in\Omega$
and a $S$-contract $c^S$ such that
$\vecc^S(\omega^*)\succ_S\vecc^N_S(\omega^*)$, where $\vecc^S(\omega^*)$ and
$\vecc^N_S(\omega^*)$ are considered as trivial payoff distributions.

In the ex-interim situation, each agent learns $j$ in which information set $I$ of their
information partition $\cI_j$ the true state of the world lies. The definition
of ex-interim blocking is quite subtle because the information partitions are
supposed to be public knowledge and we will omit its detailed definition, which
would necessitate a lengthy exposition.

No general result is given in \cite{iesh08} about the different types of core,
but simple and interesting results are given in the case where the preferences
about payoff distributions (contracts) are identical for all agents and defined
through their expected value: $X\succcurlyeq_j Y$ iff $\EE(X)\geqslant \EE(Y)$,
where expectation is understood w.r.t. the prior distribution $P$ on $\Omega$
(this is called {\it risk-neutral} preference). In that case, the ex-ante core is
characterized as follows.
\begin{theorem}\label{th:iesh}
Given a Bayesian game $(N,\Omega,P,(\cI_j)_{j\in N},(\succcurlyeq_j)_{j\in N})$
where all agents are risk-neutral,   a grand contract $\vecc^N$ belongs to the
ex-ante core iff for all $S\subseteq N$,
\[
\EE(v_\omega(S))\leqslant \sum_{j\in S}\EE(\vecc^N_j(\omega)).
\]
\end{theorem}
Recall that in the ex-ante case, the information partitions are not
used. Observe that this result is very close to the classical definition of the
core for TU-games.

For the ex-post core, the following characterization is obtained:
\begin{theorem}
Given a Bayesian game $(N,\Omega,P,(\cI_j)_{j\in N},(\succcurlyeq_j)_{j\in N})$
where all agents are risk-neutral,   a grand contract $\vecc^N$ belongs to the
ex-post core iff for all $S\subseteq N$ and for all $\omega\in\Omega$ it holds
\[
v_\omega(S) \leqslant \sum_{j\in S}\vecc^N_j(\omega).
\]
\end{theorem}
Clearly, this implies that the ex-post core is included in the ex-ante core. We
note, however, that the definition of the ex-post core is not very
natural. Indeed, usually ex-post means ``when uncertainty has been resolved'',
i.e., the true state $\omega^*$ is revealed, so that the ex-post core should be
simply the classical core of the game $v_{\omega^*}$. With this definition, the
above inclusion would not hold in general.

\section{Transferable utility games with uncertainty}\label{sec:hahe}
Pursuing our exploration of the literature, we describe now the model of Habis
and Herings \cite{hahe11} published in 2011. This model departs from all
previous ones we have described, in the sense that all of them, putting apart
some aspects which are less developed, are mainly centered on {\it ex-ante}
situations: players express preferences on uncertain payoffs before uncertainty
is resolved, and they are supposed to stick to their preference once uncertainty
is resolved, and accept the payoff which is obtained, possibly up to some small adjustment.

Habis and Herings take a different pathway. They suppose that no binding
agreement can be made ex-ante, and the players must instead discuss agreements that are
self-enforcing once uncertainty is resolved. Their definition of the core is
based on the weak sequential core, a notion proposed by Kranich et
al. \cite{krpepe05} in the context of dynamic TU-games (see also \cite{hahe10}).

We introduce formally the model. The ex-ante situation is called period or state
0, while period 1 means that uncertainty is resolved. $S$ is the set of states
of the world, and $S':=S\cup\{0\}$. For every state $s\in S$, a TU-game $v_s$ is
defined on the set $N$ of players. A utility function $u^i$ for player $i$ is defined
for evaluating uncertain payoffs $x^i\in\RR^S$ by a real number $u^i(x^i)$, and
has the following state-separable form:
\[
u^i(x^i) = \sum_{s\in S}u^i_s(x^i_s),
\]
where $u^i_s$ is increasing. Then, a {\it TU-game with uncertainty (TUU-game)}
is a tuple $\Gamma=(N,S,v,u)$ with $v=(v_s)_{s\in S}$ and $u=(u_i)_{i\in N}$. An
{\it allocation} is a distribution of the worth $v_s(N)$ for every $s\in S$,
under the form of a matrix $x=(x^1,\ldots,x^n)$ with $x^i\in\RR^S$ for all $i\in
N$. A row of this matrix is denoted by $x_s=(x_s^1,\ldots, x_s^n)$, $s\in S$
(allocation in state $s$), and $x_s(N)=v_s(N)$. Given a coalition $C\subseteq
N$, one can define $x^C$ the restriction of $x$ to $C$, as well as the
restricted game $(\Gamma,C)$.

\medskip

We turn now to the definition of the Weak Sequential Core. Let us recall first
the usual way in which agreements are done among players in the previously seen
models: agents fully commit to any state-contingent allocation (called
explicitly contracts in the model of Ieong and Shoham), implying that the set of
feasible allocation/contracts for a coalition $C$ is
$X^C=\{x^C\in\RR^{S\times C}\mymid x^C(S)\leqslant v(C)\}$, where $\leqslant$ is
understood coordinatewise. Then, by use of the utility functions $u^i$, $i\in N$, this defines for
each coalition $C$ a set of possible utility vectors:
\[
V(C):=\{\bar{u}^C\in\RR^C\mymid \exists x^C\in X^C, \forall i\in C, \bar{u}^i=u^i(x^i)\}.
\]
This defines a NTU-game, and solutions defined for NTU-games can be used to
define solutions for TUU-games.

Habis and Herings say that this construction is not realistic because agents
may deviate after learning the true state of the world. Instead, they look for
self-enforcing agreements, by using the notion of weak sequential core.

Fix some allocation $\bar{x}$. An allocation $x^C$ is {\it feasible} for
coalition $C$ at state $s\in S$ given $\bar{x}$ if $x^C$ is equal to $\bar{x}^C$ for
all states except $s$, and $x_s^C(C)\leqslant v_s(C)$.  Now, $x^C$ is {\it
  feasible at state 0} if $x^C(C)\leqslant v(C)$ (coordinatewise). Next, we define deviations. Given an allocation
$\bar{x}$, a coalition $C$ can {\it deviate} from $\bar{x}$ at $s'\in S'$ if
there exists a feasible $x^C$ at $s'$ given $\bar{x}$ (called the {\it
  deviation}) such that $u^i(x^i)>u^i(\bar{x}^i)$ for all $i\in C$,   

At this stage, observe that feasibility at state 0 is identical to feasibility
as defined by Ieong and Shoham, and deviation at state 0 is nothing other than
ex-ante blocking, see Section~\ref{sec:iesh}.

However, it can be observed on simple examples that deviations may not be
self-enforcing, in the sense that once the true state of the world is revealed,
some coalition may block the allocation $x$. This is why Habis and Herings
define credible deviations in a recursive manner. Fix some allocation
$\bar{x}$. Any deviation $x^C$ from $\bar{x}$ at $s\in S$ is {\it credible} if
$C$ is a singleton. Suppose now that credible deviations have been defined for
each coalition of size at most $k$. Let $C$, with $|C|=k+1$. A deviation $x^C$
from $\bar{x}$ at $s\in S$ is {\it credible} if there is no $D\subsetneq C$ such
that $D$ has a credible deviation from $x^C$ at $s$.

We define now credible deviations at state 0. Consider a fixed $\bar{x}$. Any
deviation $x^C$ from $\bar{x}$ at state 0 by a singleton is credible. Suppose
thet credible deviations have been defined for coalitions of size at most
$k$. Let $C$ such that $|C|=k+1$. A deviation $x^C$ from $\bar{x}$ at state 0 by
$C$ is credible id there does not exist $D\subsetneq C$ and state $s'\in S'$
such that $D$ has a credible deviation from $x^C$ at $s'$.

The {\it Weak Sequential Core} $WSC(\Gamma)$ of a TUU-game $\Gamma$ is the set
of feasible allocations $\bar{x}$ for $N$ from which no coalition has a credible
deviation.

The following result characterizes the Weak Sequential Core.
\begin{theorem}
Let $\Gamma$ be a TUU-game.  
An allocation $\bar{x}$ belongs to $WSC(\Gamma)$ if and only if $\bar{x}_s\in
\cC(N,\Gamma_s)$ for all states $s\in S$, and there is no coalition $C$ and
allocation $x^C$ such that $x^C_s\in \cC(C,\Gamma_s)$ for all $s\in S$ and
$u^i(x^i)>u^i(\bar{x}^i)$ for all $i\in C$.
\end{theorem}
The theorem says that for an allocation of belong to the Weak Sequential Core,
it should belong to the classical core of $\Gamma_s$ for every state, and no
coalition $C$ should be able to find a core element of the subgame $\Gamma_s$ restricted to
$C$ which offers better ex-ante utility for all players in $C$.

Finally, it is shown that if all games $\Gamma_s$ are convex, then the Weak
Sequential Core is nonempty.

\section{Bel coalitional games}\label{sec:bel}
Lastly, we present our own contribution to the field. Our departure point is the
notion of Bayesian coalitional game of Ieong and Shoham. Our aim is to offer
more flexibility to the model, while simplifying some of its aspects. The main
differences between Bayesian coalitional games and our notion called {\it Bel
  coalitional games} \cite{grlo24,grlo25,grlo26} are summarized below:
\begin{itemize}
\item We do not assume a common prior to all agents and we do not suppose it is
  public knowldege.
\item The prior is not necessarily a probability distribution but the more
  general notion of belief function (see below).
\item We do not consider information partitions and suppose that every agent
  can distinguish every state of the world.
\item Preference among contracts (payoff distributions) are expressed via the
  Choquet integral. This is the natural generalization of expected value when
  probability measures are replaced by belief functions and other ``non-additive
  probabilities''. 
\item In the ex-post situation, we consider that the true state of the world
  $\omega^*$ is revealed, and the notion of blocking is defined for the game
  $v_{\omega^*}$, so that we are back to the classical deterministic case.
\item Due to the disparition of information partitions, the ex-iterim situation
  has been modified. We consider that we have a decreasing sequence of
  ``information sets'' $I_1\supseteq \cdots I_t\supseteq \cdots \supseteq I_T$
  where the true state $\omega^*$ belongs to all of them, and finally
  $I_T=\{\omega^*\}$, i.e., the uncertainty is resolved step by step. 
\end{itemize}

We start by introducing the necessary background on belief functions. Full
details can be found in \cite{gra16}.
Let $\Omega = \{\omega_1,\omega_2,\ldots,\omega_d\}$ be a finite nonempty set
of states of the world.  A \textit{capacity} \cite{cho53} is a set function $\nu:2^\Omega\rightarrow [0,1]$ satisfying:
\begin{itemize}
\item[\it (i)] $\nu(\emptyset)= 0$ and $\nu(\Omega)= 1$;

\item[\it (ii)] $\nu(A)\le\nu(B)$, whenever $A \subseteq B$.
\end{itemize}
A \textit{belief function} (see \cite{dem67,sha76}) is a particular capacity
satisfying {\it total monotonicity}:
\begin{equation}\label{eq:tm}
\nu\left(\bigcup^k_{i=1} E_i\right)\ge \sum_{\emptyset \ne I \subseteq
  \{1,\ldots,k\}} (-1)^{|I|+1} \nu\left(\bigcap_{i\in I} E_i\right),
\end{equation}
for all $ k\ge 2$ and $E_1,\ldots,E_k\in 2^\Omega$. Note that convexity is
implied by total monotonicity, and that a probability measure $\pi$ on $2^\Omega$ is a particular belief
function where (\ref{eq:tm}) holds with equality. 

A belief function $\nu$ is associated with a dual set function $\overline{\nu}$ on $2^\Omega$ called \textit{plausibility function} and defined, for all $A\in 2^\Omega$, as $\overline{\nu}(A) =1-\nu(A^c).$
Both belief and plausibility functions are completely characterized by their
\textit{M\"obius inverse} (or \textit{mass function}) $m_\nu: 2^\Omega
\rightarrow [0,1]$ via the formulas
\[
\nu(A)=\sum_{B\subseteq A} m_\nu(B)\quad\mbox{and}\quad
\overline{\nu}(A)=\sum_{B\cap A \neq \emptyset} m_\nu(B).
\]
A subset of $\Omega$ with strictly positive mass function is called a {\it focal element} and we denote by $\mathcal{F}_{\nu} = \{E \in 2^\Omega \,:\, m_\nu(E) > 0\}$ the set of focal elements of $\nu$.

Given a capacity $\nu$ and $X \in \mathbb{R}^\Omega$, the {\it Choquet expectation} of $X$ with respect to $\nu$ is defined through the Choquet
integral \cite{cho53}:
$$\mathbb{C}_{\nu}(X):=\int X \,\d\nu =\sum^d_{i=1}[X(\omega_{\sigma(i)})-X(\omega_{\sigma(i+1)})]\nu(E_i^\sigma),$$ 
where $\sigma$ is a permutation of $\{1,\ldots,d\}$ such that
$X(\omega_{\sigma(1)})\ge \cdots \ge X(\omega_{\sigma(d)})$,
$E_i^\sigma=\{\omega_{\sigma(1)},\ldots,\omega_{\sigma(i)}\}$, for
$i=1,\ldots,d$, and $X(\omega_{\sigma(d+1)})=0$. When $\nu$ is a
probability measure $\pi$, we have that $\mathbb{C}_\pi(X)=\mathbb{E}_\pi(X)$,
the expected value of $X$. 

We recall that classical expectation w.r.t. a probability measure induces (or is
defined as) risk neutral attitude. Interestingly, the Choquet expectation
w.r.t. a convex capacity (and hence a belief function) induces a risk loving
attitude, while a concave capacity (e.g., a plausibility function) induces a
risk averse attitude. Moreover, the Choquet integral is
superadditive/subadditive for convex/concave capacities, implying that if $\nu$
is a belief function we have for every $X,Y\in\RR^\Omega$:
\[
\int(X+Y)\d\nu\geqslant \int X\d\nu+\int Y\d\nu, \quad \int(X+Y)\d\overline{\nu}\leqslant \int X\d\overline{\nu}+\int Y\d\overline{\nu}.
\]
This ability of modelling risk attitude is one of the main motivation of using
belief functions. Moreover, they can adequately represent ignorance, and are
able to distinguish equal certainty of the states of the world from total
ignorance.

\medskip

We now turn to the description of our setting. 
\begin{definition}
    A \textit{Bel coalitional game} is given by $(N,\Omega,(m_{\nu_j})_{j\in N},(v_\omega)_{\omega\in\Omega})$ where
    \begin{itemize}
        \item $N= \{1,2,\ldots,n\}$ is a set of agents;
        \item $\Omega= \{\omega_1,\omega_2,\ldots,\omega_d\}$ is a set of
          possible worlds, where each world $\omega$ specifies a coalitional game $v_\omega$ defined over $N$;
         \item $m_{\nu_j}:2^\Omega\rightarrow [0,1]$ is agents $j$'s mass distribution over the worlds in $\Omega$, and $\nu_j$ is the corresponding belief function, such that $\bigcup\mathcal{F}_{\nu_j}=\Omega$ (the
             focal elements cover $\Omega$);
       \item $v_\omega: 2^N \rightarrow \mathbb{R}$ is a coalitional game that assigns to each coalition $S\subseteq N$ its value in world $\omega$.
    \end{itemize}
\end{definition}

For the sake of brevity, we focus in what follows on the ex-ante
situation. Indeed, the ex-interim situation does not differ so much from the
ex-ante situation, while as we said above, the ex-post boils down to the
classical deterministic case. 

We borrow without any change the notion of {\it contract} and {\it feasible
  contract} from Ieong and Shoham.  We say that a (feasible) grand contract $\textbf{c}^N$ is {\it efficient at $\omega$} if $\sum_{j\in
  N}\textbf{c}^N_j(\omega)=v_\omega(N)$. Therefore, it is not efficient if $\sum_{j\in
  N}\textbf{c}^N_j(\omega)<v_\omega(N)$.
We say that $\textbf{c}^N$ is {\it efficient} if it is efficient at every
$\omega\in\Omega$.

The next step is to define preference over contracts. Let
  $\vecc^S,\vecc'^S$ be two contracts for $S$. We have that $\vecc^S\succcurlyeq_S
  \vecc'^S$ if every agent $j$ in $S$ prefers the payoff distribution $\vecc^S_j$ to $\vecc'^S_j$, which we
  denote by $\vecc^S_j\succcurlyeq_j\vecc'^S_j$. We write  $\vecc^S\succ_S
  \vecc'^S$ if the preference is strict for each agent in $S$.  As explained above,
  we consider that preference over payoff distributions is given by
  the Choquet integral, i.e., for any agent $j\in S$: 

\[
\vecc^S_j\succcurlyeq_j\vecc'^S_j \Leftrightarrow \mathbb{C}_{\nu_j}(\textbf{c}_j^{S})\ge\mathbb{C}_{\nu_j}({\textbf{c}'}_j^{S}).
\]

Preferences being defined, we can borrow without change the notion of ex-ante
blocking from Ieong and Shoham, and similarly, we define the {\it ex-ante core}
as the set of feasible grand contracts which are not ex-ante blocked by any coalition.  

The properties of the ex-ante core are explored at length in
\cite{grlo25}. However, in the general case where priors are belief functions,
not necessarily identical for all agents, no characterization result is
obtained. Instead, a necessary condition is obtained.

\begin{theorem}\label{th1}
Given a Bel coalitional game $(N,\Omega,(m_{\nu_j})_{j\in N},(v_\omega)_{\omega\in\Omega})$, if a grand contract $\textbf{c}^N$ is in the ex-ante core of the game then, for all $S\subseteq N$, there exists $j\in S$ such that
$$\mathbb{C}_{\nu_j}(v_\omega(S))\le\mathbb{C}_{\overline{\nu}_j}\left(\sum_{i\in S}\textbf{c}^N_i\right).$$
\end{theorem}

To get the converse property, one needs to suppose that all priors are
identical.
\begin{theorem}\label{th2}
Let us consider that agents have the same belief function $\nu$. Given a Bel
coalitional game $(N,\Omega,m_{\nu},(v_\omega)_{\omega\in\Omega})$ and a grand contract $\textbf{c}^N$, if for all $S\subseteq N$
$$\mathbb{C}_{\nu}(v_\omega(S))\le\sum_{j\in S}\mathbb{C}_{\nu}(\textbf{c}^N_j),$$
then the grand contract $\textbf{c}^N$ is in the ex-ante core of the game.
\end{theorem}
Observe that when all priors are probability measures, then the result of Ieong
and Shoham (Theorem~\ref{th:iesh}) is recovered.

In \cite{grlo26}, thanks to the characterization of the ex-ante core when all
priors are identical probability measures, the geometrical structure of the
ex-ante core is elucidated. The first result
says that the ex-ante core is an affine space and give a basis of the
corresponding vector space, of dimension $(n-1)(d-1)$. Denoting
$\pi=(\pi_1,\ldots,\pi_d)$ the common prior, the basis is
$(\vecw^1_2,\ldots,\vecw^1_n,\vecw^2_2,\ldots,\vecw^2_n,\ldots, \vecw^{d-1}_2,\ldots,\vecw^{d-1}_n)$,
given by
\begin{align*}
  \vecw^1_2 &=
  \Big(\underbrace{-1,1,0,\ldots,0}_{j=1},0\ldots,0,\underbrace{\frac{\pi_1}{\pi_d},-\frac{\pi_1}{\pi_d},0,\ldots,0}_{j=d}\Big)\\
  \vecw^1_n &=
  \Big(\underbrace{-1,0,\ldots,0,1}_{j=1},0\ldots,0,\underbrace{\frac{\pi_1}{\pi_d},0,\ldots,0,-\frac{\pi_1}{\pi_d}}_{j=d}\Big)\\
    \vecw^2_2 &=
  \Big(0,\ldots,0,\underbrace{-1,1,0,\ldots,0}_{j=2},0\ldots,0,\underbrace{\frac{\pi_2}{\pi_d},-\frac{\pi_2}{\pi_d},0,\ldots,0}_{j=d}\Big)\\
  \vecw^2_n &=
  \Big(0,\ldots,0,\underbrace{-1,0,\ldots,0,1}_{j=2},0\ldots,0,\underbrace{\frac{\pi_2}{\pi_d},0,\ldots,0,-\frac{\pi_2}{\pi_d}}_{j=d}\Big)\\
  \vecw^{d-1}_2 &=
  \Big(0,\ldots,0,\underbrace{-1,1,0,\ldots,0}_{j=d-1},\underbrace{\frac{\pi_{d-1}}{\pi_d},-\frac{\pi_{d-1}}{\pi_d},0,\ldots,0}_{j=d}\Big)\\
  \vecw^{d-1}_n &=
  \Big(0,\ldots,0,\underbrace{-1,0,\ldots,0,1}_{j=d-1},\underbrace{\frac{\pi_{d-1}}{\pi_d},0,\ldots,0,-\frac{\pi_{d-1}}{\pi_d}}_{j=d}\Big)
\end{align*}

To get further insight, we introduce the expected TU-game $V$ on $N$ defined by
\[
V(S) =\sum_{\omega\in\Omega}\pi(\omega)v_{\omega}(S), \quad (S\subseteq N).
\]
\begin{theorem}
Consider a balanced Bel coalitional game $(N,\Omega,m_\nu,(v_\omega)_{\omega\in\Omega})$
with $\nu=\pi$. The following holds.
\begin{enumerate}
\item $\cC((v_\omega)_{\omega\in\Omega},\pi)\neq\emptyset$ iff
  $\cC(V)\neq\emptyset$;
\item Supposing $\cC(V ) \neq \emptyset$, to any vertex $y \in \RR^N$ of $\cC(V
  )$ corresponds a pseudo-vertex $x=(x^1,\ldots,x^d)$ in $\RR^{N \times\Omega}$ of
  $\cC((v_\omega)_{\omega\in\Omega},\pi)$ and conversely, where the correspondance
  is given by:
  \begin{align*}
    &\sum_{j=1}^d\pi_jx^j = y\\
    & x^j(N)= v_{\omega_j}(N), \quad j=1,\ldots, d.
  \end{align*}
\end{enumerate}
\end{theorem}

We turn to the definition of convex games. A Bel coalitional game
$(N,\Omega,m_{\nu},(v_\omega)_{\omega\in\Omega})$ with identical priors is {\it ex-ante convex} if for all $S,T\subseteq N$
$$\mathbb{C}_{\overline{\nu}}(v_\omega(S))+\mathbb{C}_{\overline{\nu}}(v_\omega(T))\le \mathbb{C}_{\nu}(v_\omega(S\cup T))+\mathbb{C}_{\nu}(v_\omega(S\cap T)).$$

The question is whether convexity implies nonemptiness of the ex-ante core, and
even more, if the marginal vectors belong to it as in the classical case. When
all priors are identical and a probability distribution $\pi$, the answer is
positive, and immediate from the above results, because ex-ante convexity in
this case is nothing other than convexity of the expected game $V$. The next
result shows that this is still true if the common prior is a belief function
$\nu$. For each $v_\omega$ and permutation $\sigma$ on $N$, the marginal vector
$m^{\sigma,v_\omega}$ is the classical one, and we consider for each permutation
$\sigma$ the {\it marginal contract}
$(m^{\sigma,v_\omega})_{\omega\in\Omega})$. Then, the following holds.

\begin{theorem}\label{thconvex1}
Let us consider a Bel coalitional game
$(N,\Omega,m_{\nu},(v_\omega)_{\omega\in\Omega})$. If it is ex-ante convex, then
$(m^{\sigma,v_\omega})_{\omega\in\Omega}$ is in the ex-ante core of the game for
all permutations $\sigma$ on $N$.
\end{theorem}

\medskip

In \cite{grlo26}, other solution concepts are defined in this framework, namely
the (pre-)nucleolus, the kernel and the bargaining set. For the sake of conciseness,
we mention very briefly the nucleolus for comparison with the framework of
Charnes and Granot, and give some insights in the bargaining
set.

The key point to define the nucleolus is the notion of excess. We define the {\it ex-ante excess} of some coalition $S$
to a grand contract $\textbf{c}^N$ as
$$e(S,\textbf{c}^N)=\mathbb{C}_{\nu}(v_\omega(S))-\sum_{i\in S}\mathbb{C}_{\nu}(\textbf{c}^N_i).$$
As this is a real quantity, the rest follows the classical way: the nucleolus
lexicographically minimizes the vector of decreasing excesses among all grand
contracts, while the pre-nucleolus does the same over efficient grand
contracts. It is proved that the pre-nucleolus always exists, but unlike the
classical case, is not necessarily a singleton.

We have explained in Section~\ref{sec:prel} that there are two main definitions
for the bargaining set, the Davis and Maschler definition and the Mas-Colell definition.
 In the definition of Mas-Colell, recall that
instead of one player having objection against another player, we have that one
coalition has an objection against another one. It appears that the definition
of an objection at $x$ by a coalition $S$ is close to our notion of blocking,
where a coalition $S$ blocks some grand contract by an $S$-contract. For this
reason we adopt the definition of Mas-Colell. 

Given a Bel coalitional game $(N,\Omega,(m_{\nu_j})_{j\in
  N},(v_\omega)_{\omega\in\Omega})$, a grand contract $\textbf{c}^N$ and an
$S$-contract $\textbf{c}^S$ such that $\textbf{c}^S\succ_S\textbf{c}_S^N$n
(i.e., $S$ blocks $\vecc^N$),
a coalition $S'$ {\it ex-ante counterblocks} $\textbf{c}^S$ if there exists a feasible $S'$-contract $\textbf{c}^{S'}$ such that
\begin{itemize}
\item $\mathbb{C}_{\nu_j}(\textbf{c}^{S'}_j)\ge\mathbb{C}_{\nu_j}(\textbf{c}^S_j)$, for all $j\in S'\cap S$
\item $\mathbb{C}_{\nu_j}(\textbf{c}^{S'}_j)\ge\mathbb{C}_{\nu_j}(\textbf{c}_j^N)$, for all $j\in S'\setminus S$
\end{itemize}
and at least one of these inequalities is strict. Now, we say that a coalition
$S$ {\it legitimate ex-ante blocks} a grand contract $\textbf{c}^N$ if there
does not exist any counterblocking coalition $S'$ to the contract
$\textbf{c}^{S}$. The {\it ex-ante bargaining set} of a Bel coalitional game is
the set of all feasible grand contracts $\textbf{c}^N$ for which no coalition
$S\subseteq N$ legitimate ex-ante blocks $\textbf{c}^N$.

In \cite{grlo26}, we show that the ex-ante bargaining set always exist, however,
unlike the classical case, it may be not a compact set (as the ex-ante core).

The last question we address in this survey is about convex games. It is
well-known that for classical cooperative games, the bargaining set in the sense
of Davis and Maschler coincides with the core for convex games. Up to our
knowedge, this property has not been proved for the bargaining set in the sense
of Mas-Colell.
What we propose below is a strengthening of our original definition,
resulting in a smaller bargaining set, which
makes true the coincidence of the core and the bargaining set for convex
games. In a sense, it is closer to the definition of Davis and Maschler, as we
will explain.

The {\it ex-ante strong bargaining set} of a Bel coalitional game is the set of all feasible grand contracts $\textbf{c}^N$ for which every ex-ante blocking coalition $S\subseteq N$ admits a counterblocking coalition $S'$ to the contract $\textbf{c}^{S}$ such that $S'\not\subseteq S$.

Recall that in the definition of Davis and Maschler, objections
  are defined w.r.t. players: $i$ has an objection against $j$ if there is a
  coalition $S$ containing $i$ but not $j$ and an imputation on $S$ making all
  players in $S$ better off. Then a counterobjection of $j$ against $i$ amounts
  to find a coalition $T$ containing $j$ but not $i$, and an imputation better
  for players in $T$. As a consequence, it can never be the case that $T$ is
  included in $S$. This is exactly what we propose: the coalition which makes a
  counterobjection cannot be a subset of the coalition making an objection.
  
\begin{theorem}\label{thcorebarg}
Let us consider a Bel coalitional game $(N,\Omega,m_{\nu},(v_\omega)_{\omega\in\Omega})$, with $\nu=\pi$. If it is ex-ante convex, then the ex-ante strong bargaining set coincide with the ex-ante core.
\end{theorem}

\section{Concluding remarks}
This survey has shown the variety of the proposed models for incorporating
uncertainty in cooperative games, and we have tried to enhance difference and
similarities between these models. Most of them, including ours, and with
the notable exception of the model of Habis and Herings, make an ex-ante
treatment of uncertainty, removing uncertainty by taking expectation (either in
the classical sense with probability measure or in a more sophisticated way with
belief functions) or quantiles (Charnes and Granot) or by mixing the expected
value with the ex-post value (Suijs and Borm). Interestingly, most authors
define the core via a blocking mechanism, going back to the notion of dominance
core, which in the literature of (deterministic) cooperative games, has been
rapidly replaced by the notion of core by Gillies \cite{gil53} based on
coalitional rationality. A third interesting aspect which emerge from the survey is the
relation with NTU-games, another facet of cooperative games fallen somehow into
oblivion. As pointed out by Habis and Herings, one could simply solve the
problem of treating cooperative TU-games with uncertainty by turning them into
NTU-games. This is however in some sense like taking a step back to leap
forward: the difficulty of dealing with NTU-games and the multiplicity of
their solution concepts is well known...

\bibliographystyle{plain}

\bibliography{../BIB/fuzzy,../BIB/grabisch,../BIB/general}

\begin{thebibliography}{10}

\bibitem{bon63}
O.~Bondareva.
\newblock Some applications of linear programming to the theory of cooperative
  games.
\newblock {\em Problemy Kibernetiki}, 10:119--139, 1963.
\newblock in Russian.

\bibitem{bodepe05}
W.~Bossert, J.~Derks, and H.~Peters.
\newblock Efficiency in uncertain cooperative games.
\newblock {\em Mathematical Social Sciences}, 50:12--23, 2005.

\bibitem{brbrgoti10}
R.~Branzei, O.~Branzei, S.~Z. {Alparslan G\"ok}, and S.~Tijs.
\newblock Cooperative interval games: a survey.
\newblock {\em Central European J. of Operations Research}, 18:397--411, 2010.

\bibitem{brditi05}
R.~Branzei, D.~Dimitrov, and S.~Tijs.
\newblock {\em Models in cooperative game theory: crisp, fuzzy and multichoice
  games}.
\newblock Springer Verlag, 2005.

\bibitem{chgr73}
A.~Charnes and D.~Granot.
\newblock Prior solutions: Extensions of convex nucleolus solutions to
  chance-constrained games.
\newblock In {\em Proc. of the Computer Science and Statistics 7th Symposium},
  pages 323--332, Iowa State University, 1973.

\bibitem{chgr76}
A.~Charnes and D.~Granot.
\newblock Coalitional and chance-constrained solutions to $n$-person games. i:
  the prior satisficing nucleolus.
\newblock {\em SIAM J. on Applied Mathematics}, 31(2):358--367, 1976.

\bibitem{chgr77}
A.~Charnes and D.~Granot.
\newblock Coalitional and chance-constrained solutions to $n$-person games. ii:
  two stages solutions.
\newblock {\em Operations Research}, 25:1013--1019, 1977.

\bibitem{cho53}
G.~Choquet.
\newblock Theory of capacities.
\newblock {\em Annales de l'Institut Fourier}, 5:131--295, 1953.

\bibitem{dama67}
M.~Davis and M.~Maschler.
\newblock Existence of stable payoff configurations for cooperative games.
\newblock In M.~Shubik, editor, {\em Essays in Mathematical Economics in Honor
  of Oskar Morgenstern}, pages 39--52. Princeton University Press, 1967.

\bibitem{dem67}
A.~P. Dempster.
\newblock Upper and lower probabilities induced by a multivalued mapping.
\newblock {\em Ann. Math. Statist.}, 38:325--339, 1967.

\bibitem{gil53}
D.~Gillies.
\newblock {\em Some theorems on $n$-person games}.
\newblock PhD thesis, Princeton, New Jersey, 1953.

\bibitem{gra12a}
M.~Grabisch.
\newblock The core of games on ordered structures and graphs.
\newblock {\em Annals of Operations Research}, 204:33--64, 2013.
\newblock doi: 10.1007/s10479-012-1265-4.

\bibitem{gra16}
M.~Grabisch.
\newblock {\em Set Functions, Games and Capacities in Decision Making},
  volume~46 of {\em Theory and Decision Library C}.
\newblock Springer, 2016.

\bibitem{grlo24}
M.~Grabisch and S.~Lorenzini.
\newblock Bel coalitional games.
\newblock In S.~Destercke, M.V. Martinez, and G.~Sanfilippo, editors, {\em 16th
  International Conference on Scalable Uncertainty Management (SUM 2024)},
  volume 15350 of {\em Lecture Notes in Computer Sciences}. Springer, 2024.

\bibitem{grlo25}
M.~Grabisch and S.~Lorenzini.
\newblock Bel coalitional games with application to market games.
\newblock {\em Int. J. of Approximate Reasoning}, 184:109466, 2025.

\bibitem{grlo26}
M.~Grabisch and S.~Lorenzini.
\newblock Properties of the core and other solution concepts of bel coalitional
  games in the ex-ante scenario.
\newblock arXiv:2602.04817, 2026.

\bibitem{gra77}
D.~Granot.
\newblock Cooperative games in stochastic characteristic function form.
\newblock {\em Management Sciences}, 23:621--630, 1977.

\bibitem{hahe10}
H.~Habis and P.J.J. Herings.
\newblock A note on the weak sequential core of dynamic {TU} games.
\newblock {\em Int. Game Theory Review}, 12:407--416, 2010.

\bibitem{hahe11}
H.~Habis and P.J.J. Herings.
\newblock Transferable utility games with uncertainty.
\newblock {\em J. of Economic Theory}, 146:2126--2139, 2011.

\bibitem{hsra93}
C.~R. Hsiao and {T. E. S.} Raghavan.
\newblock Shapley value for multichoice cooperative games, {I}.
\newblock {\em Games and Economic Behavior}, 5:240--256, 1993.

\bibitem{iesh08}
S.~Ieong and Y.~Shoham.
\newblock Bayesian coalitional games.
\newblock In {\em Proceedings of the 23d AAAI Conference on Artificial
  Intelligence}, pages 95--100, Chicago, IL, July 2008.

\bibitem{krpepe05}
L.~Kranich, A.~Perea, and H.~Peters.
\newblock Core concepts for dynamic tu games.
\newblock {\em Int. Game Theory Review}, 7:43--61, 2005.

\bibitem{mar94}
M.~Mare\v{s}.
\newblock {\em Computation over Fuzzy Quantities}.
\newblock CRC Press, 1994.

\bibitem{mas89}
A.~Mas-Colell.
\newblock An equivalence theorem for a bargaining set.
\newblock {\em J. of Mathematical Economics}, pages 129--139, 1989.

\bibitem{masoza13}
M.~Maschler, E.~Solan, and S.~Zamir.
\newblock {\em Game Theory}.
\newblock Cambridge University Press, 2013.

\bibitem{pesu03}
B.~Peleg and P.~Sudh\"olter.
\newblock {\em Introduction to the theory of cooperative games}.
\newblock Kluwer Academic Publisher, 2003.

\bibitem{pet08}
H.~Peters.
\newblock {\em Game Theory: A Multilevel Approach}.
\newblock Springer, 2008.

\bibitem{sch69}
D.~Schmeidler.
\newblock The nucleolus of a characteristic function game.
\newblock {\em SIAM J. on Applied Mathematics}, 17:1163--1170, 1969.

\bibitem{sha76}
G.~Shafer.
\newblock {\em A Mathematical Theory of Evidence}.
\newblock Princeton Univ. Press, 1976.

\bibitem{sha67}
L.~S. Shapley.
\newblock On balanced sets and cores.
\newblock {\em Naval research Logistics Quarterly}, 14:453--460, 1967.

\bibitem{subo99}
J.~Suijs and P.~Borm.
\newblock Stochastic cooperative games: Superadditivity, convexity and
  certainty equivalents.
\newblock {\em Games and Economic Behavior}, 27:331--345, 1999.

\bibitem{subowati99}
J.~Suijs, P.~Borm, A.~De Waegenaere, and S.~Tijs.
\newblock Cooperative games with stochastic payoffs.
\newblock {\em European J. of Operational Research}, 113:193--205, 1999.

\bibitem{thlu63}
R.~M. Thrall and W.~F. Lucas.
\newblock $n$-person games in partition function form.
\newblock {\em Naval Research Logistic Quarterly}, 10:281--298, 1963.

\bibitem{vnm44}
J.~{von Neumann} and O.~Morgenstern.
\newblock {\em Theory of Games and Economic Behavior}.
\newblock Princeton University Press, 2nd edition, 1947.

\end{thebibliography}

\end{document}